\documentclass{ieeeaccess}
\usepackage{lmodern}
\usepackage{cite}
\usepackage[inline, shortlabels]{enumitem}
\usepackage{mathrsfs}
\usepackage[UKenglish]{babel}
\usepackage{amsmath,amssymb,amsfonts,amsthm}
\newtheorem{lemma}{Lemma}

\newtheorem{definition}{Definition}

\newtheorem{theorem}{Theorem}
\newtheorem{observation}{Observation}
\usepackage{float}
\usepackage{algorithm}
\usepackage{algpseudocode}
\usepackage{listings}
\usepackage{graphicx}
\usepackage{textcomp}
\usepackage{mathtools}
\usepackage{centernot}
\usepackage{subcaption}
\usepackage{hyperref}

\usepackage{threeparttable}
\usepackage[noabbrev,nameinlink]{cleveref}
\usepackage{pgfplots}
\usepackage{pgfplotstable}
\usepackage{tikz}
\usepackage[outdir=fig/]{epstopdf}
\usepackage[hyperref=true]{acro}
\usepackage{color}
\usepackage{here}
\DeclareAcronym{GALS}{
	short=GALS,
	long=Globally Asynchronous Locally Synchronous
}
\DeclareAcronym{LSN}{
	short=LSN,
	long=Logical Synchrony Network,
	short-indefinite=an,
	plural=s
}
\DeclareAcronym{KPN}{
	short=KPN,
	long=Kahn Process Network
}
\DeclareAcronym{FFP}{
	short=FFP,
	long=Finite FIFO Platform,
	short-indefinite=an
}
\DeclareAcronym{LSFP}{
	short=LSFP,
	long=Logically Synchronous FIFO Platform,
	short-indefinite=an
}
\DeclareAcronym{SDFG}{
	short=SDFG,
	long=Synchronous Dataflow Graph,
	short-indefinite=an
}
\DeclareAcronym{LTTA}{
	short=LTTA,
	long=Loosely Time Triggered Architecture,
	short-indefinite=an
}
\DeclareAcronym{SBD}{
	short=SBD,
	long=Synchronous Block Diagram,
	short-indefinite=an
}
\DeclareAcronym{CPO}{
	short=CPO,
	long=Complete Partial Order
}
\DeclareAcronym{FIFO}{
	short=FIFO,
	long=First In First Out
}
\DeclareAcronym{SDF}{
	short=SDF,
	long=Synchronous Data Flow,
	short-indefinite=an
}
\DeclareAcronym{MDG}{
	short=MDG,
	long=Marked Directed Graph,
	short-indefinite=an
}
\DeclareAcronym{MoC}{
	short=MoC,
	long=Model of Computation,
	long-plural-form=Models of Computation,
	short-indefinite=an
}
\DeclareAcronym{sLET}{
	short=sLET,
	long=synchronous Logical Execution Time,
	short-indefinite=an
}
\DeclareAcronym{TTA}{
	short=TTA,
	long=Time Triggered Architecture
}

\NewSpotColorSpace{PANTONE}
\AddSpotColor{PANTONE} {PANTONE3015C} {PANTONE\SpotSpace 3015\SpotSpace C}{1 0.3 0 0.2}
\SetPageColorSpace{PANTONE}
\Crefname{observation}{observation}{observations}
\Crefname{observation}{Observation}{Observations}

\DeclareSymbolFont{matha}{OML}{txmi}{m}{it}% txfonts
\DeclareMathSymbol{\varv}{\mathord}{matha}{118}

\def\BibTeX{{\rm B\kern-.05em{\sc i\kern-.025em b}\kern-.08em
    T\kern-.1667em\lower.7ex\hbox{E}\kern-.125emX}}

\newcounter{edremcounter}

\newcommand{\ignore}[1]{{}}

\definecolor{highlight}{RGB}{0,0,255}
\definecolor{highlightg}{RGB}{0,128,0}

\newcommand{\highlight}[1]{#1}
\newcommand{\highlightg}[1]{#1}

\pgfplotsset{compat=1.17}
\theoremstyle{definition}

\newtheorem{example}{Example}
\begin{document}
\history{}
\doi{10.1109/ACCESS.2024.3411017}

\title{Logical Synchrony Networks: A formal model for deterministic distribution}
\author{\uppercase{Logan Kenwright}\authorrefmark{1},
    \uppercase{Partha Roop}\authorrefmark{1}, \uppercase{Nathan Allen}\authorrefmark{1}, \uppercase{Sanjay Lall}\authorrefmark{2}, \uppercase{C\u{a}lin Ca\c{s}caval}\authorrefmark{3},  \uppercase{Tammo Spalink}\authorrefmark{3}, \uppercase{Martin Izzard}\authorrefmark{3}}
\address[1]{Department of Electrical, Computer and Software Engineering, The University of Auckland, New Zealand}
\address[2]{Department of Electrical Engineering at Stanford University, Stanford, CA 94305, USA, and visiting researcher at Google}
\address[3]{Google Research}

\markboth
{Kenwright \headeretal: Logical Synchrony Networks: A formal model for deterministic distribution}
{Kenwright \headeretal: Logical Synchrony Networks: A formal model for deterministic distribution}
\corresp{Corresponding author: Logan Kenwright (e-mail: logan.kenwright@auckland.ac.nz).}

%\corresp{Corresponding author: First A. Author (e-mail: author@ boulder.nist.gov).}
\begin{abstract}

%%Revised abstract
\highlight{In the modelling of distributed systems, most \acp{MoC} rely on blocking communication to preserve determinism. A prominent example is} \Acp{KPN}, which supports non-blocking writes and blocking reads, and its implementable variant \acp{FFP} which enforces boundedness using blocking writes. 
An issue with these models is that they mix process synchronisation with process execution, necessitating frequent blocking during synchronisation. This paper explores a recent alternative called \emph{bittide},
which decouples the execution of a process
from the synchronisation behaviour. Determinism and boundedness is preserved while enabling pipelined execution for better throughput.
To understand the behaviour of these systems we define a formal model -- a deterministic \ac{MoC} called \acp{LSN}. \Ac{LSN}s describes
a network of processes modelled as a graph, with edges representing
\emph{invariant logical delays} between a producer process and the corresponding
consumer process. We show that this abstraction is satisfied by the \ac{KPN} model, and subsequently by both the concrete \acp{FFP} and bittide architectures.
Thus, we show that \acp{FFP} and bittide offer two ways of implementing deterministic distributed systems, with the latter being more performant.

\end{abstract}


\titlepgskip=-15pt
\maketitle

\acresetall
\acuse{FIFO}

\section{Introduction}
\label{sec:introduction}
%\PARstart{A}

\begin{table*}[htbp]
    \caption{Communication Schemes for Synchronisation}
    \centering
    \begin{tabular}{|m{2.6cm}|m{0.9cm}|m{0.9cm}|m{2.1cm}|m{1.1cm}|m{1.1cm}|m{1.1cm}|m{2cm}|m{1.6cm}|}
        \hline
        Scheme                                 & \centering Writers & \centering Readers & Buffer Size   & Blocking Read & Blocking Write & Single reads & Triggering & Decoupled Execution \\
        \hline
        Unsynchronized                         & Many               & Many               & One           & No            & No             & No & None       & No         \\
        Read-Modify-Write                      & Many               & Many               & One           & Yes           & Yes           & No & Event       & No         \\
        Unbounded \acs*{FIFO}                  & Single             & Single             & Unbounded     & Yes           & No             & Yes & Event       & No         \\
        Bounded \acs*{FIFO}                    & Single             & Single             & Bounded       & Yes           & Yes            & Yes & Event/Schedule       & No         \\
        Rendezvous                             &
        Single                                 & Varies             & One                & Yes           & Varies        & Yes & Event             & No                    \\
        Time Triggered                         & Single             & Varies             & One/Bus-based & No            & No             & Yes & Physical Clock & No         \\
        \hline
        Elastic Buffer~\cite{lall_logicalsync} & Single             & Single             & Bounded       & No            & No             & Yes & Logical Clock  & Yes        \\
        \hline
    \end{tabular}
    \label{table:mechanisms}
\end{table*}

\highlight{The modelling of distributed systems relies heavily on } \Acp{MoC}~\cite{lee1998framework}, which are used to formally describe how concurrent components of a system are composed, focussing on their computation and the communication. Many \acp{MoC} have been developed, which range from non-deterministic
models such as process algebras~\cite{hoare1985communicating, milner1989communication} and actor-based models~\cite{agha1986actors} to deterministic variants based on \acp{KPN}~\cite{gilles1974semantics} and synchronous models~\cite{benveniste_synchronous_2003}. 
\highlight{In distributed systems, the communication scheme between processes is a defining feature of the chosen \ac{MoC}.} Edwards et al.~\cite{edwards1997design} categorize common communication schemes, which we summarise in \Cref{table:mechanisms}.

\highlightg{The characteristics of these schemes include the number of writers and readers, buffer size, blocking I/O requirements, and whether a piece of data can be read at most once during a communication. Additionally, we introduce some metrics not present in~\cite{edwards1997design}: triggering, referring to the mechanism driving each communication event, and decoupled execution, denoting schemes where synchronization behavior does not interfere with process execution. A summary of common schemes is as follows:
\begin{itemize}
    \item \textit{Unsynchronized}: No coordination between send/receive, offering no guarantee on data arrival.
    \item \textit{Read-Modify-Write:} Processes communicate over shared memory, employing locking mechanisms.
    \item \textit{Unbounded \ac{FIFO}:} Sender-generated data tokens are consumed by receivers in order, allowing any number of tokens in the intermediate buffer.
    \item \textit{Bounded \ac{FIFO}:} Similar to unbounded FIFO, but with a finite buffer capacity. A sender cannot progress without free space available.
    \item \textit{Rendezvous:} Processes synchronize through explicit messaging, potentially causing blocking until matching read/write events occur.
\end{itemize}}

\highlightg{In addition, we note the following schemes which have been developed since~\cite{edwards1997design}:
\begin{itemize}
    \item \textit{Time triggered:} Synchronization based on known timing bounds, utilizing either physical or logical clocks.~\cite{maier_time-triggered_2002}.
    \item \textit{Elastic Buffer:} A novel mechanism found in bittide~\cite{lall_modeling_2021}.
    Provides bounded FIFO behaviour without blocking reads or writes, by varying communication speed using a control mechanism.
\end{itemize}}

\highlightg{Among these schemes, deterministic models, particularly FIFO-based models, hold significance for designing safety-critical systems.} Kahn's seminal work formalises \ac{KPN}
models~\cite{gilles1974semantics}, where a producer never blocks, while a consumer
blocks while accessing an empty \ac{FIFO}.
When a consumer blocks on a given \ac{FIFO}, it is prevented from
context switching and hence this model is shown to be \emph{determinate}~\cite{gilles1974semantics},
i.e. for any arbitrary
execution order
of the processes, the order of tokens in the buffers remains invariant.
However, the scheduling problem using bounded memory is undecidable. Hence,
variants such as \acp{FFP}~\cite{tripakis_implementing_2008}
introduce blocking on both the
producer side (when the \ac{FIFO} is full) and consumer side
(when the \ac{FIFO} is empty). We observe that for synchronisation, existing deterministic models either use physical time or intertwine process execution with synchronisation, leading to undesirable blocking inefficiencies. How can this blocking time be mitigated?

\subsection{\highlight{Motivation}}

\begin{figure}[!htbp]
    \centering
    \begin{subfigure}[t]{\linewidth}
        \centering
    \includegraphics[clip, trim=0.5cm 0.2cm 0.5cm 0.2cm,width=0.7\linewidth]{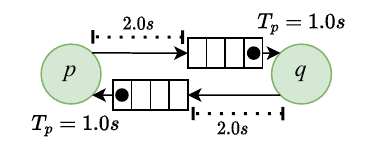}
    \caption{Tokens available in the buffers, machines are live (green)}
    \label[figure]{fig:delay_1_a}
    \end{subfigure}
    \begin{subfigure}[t]{\linewidth}
        \centering
        \includegraphics[clip, trim=0.5cm 0.2cm 0.5cm 0.2cm,width=0.7\linewidth]{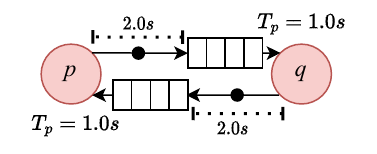}
        \caption{Tokens are in-flight, machines are not live (red)}
        \label[figure]{fig:delay_1_b}
    \end{subfigure}
    \caption{A simple FFP. Each machine has an execution time of 1 second and each link a delay of 2.0 seconds}
    \label[figure]{fig:delay_1}
\end{figure}

\highlight{Consider two machines, shown as circles in \Cref{fig:delay_1_a}, which communicate over a bounded \ac{FIFO}, shown as boxes. In order to perform a cycle of execution, each machine must consume a token from its input buffer, and produce a token on the output link. Each machine has an execution time of 1 second, meaning that each machine can nominally execute once per second. However, these two machines are separated by a lengthy transmission delay of 2 seconds, and have only a single initial token in each buffer. Consequently, more time is spent waiting for an in-flight token to arrive (\Cref{fig:delay_1_b}) than the time spent doing useful execution. Thus, the effective execution time of both machines is 3.0 seconds.}

\begin{figure}[!htbp]
    \centering
    \begin{subfigure}[t]{\linewidth}
        \centering
    \includegraphics[clip, trim=0.5cm 0.2cm 0.5cm 0.2cm,width=0.7\linewidth]{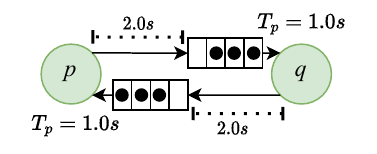}
    \caption{Tokens are available in the buffers, machines are live}
    \label[figure]{fig:delay_3_a}
    \end{subfigure}
    \begin{subfigure}[t]{\linewidth}
        \centering
        \includegraphics[clip, trim=0.5cm 0.2cm 0.5cm 0.2cm,width=0.7\linewidth]{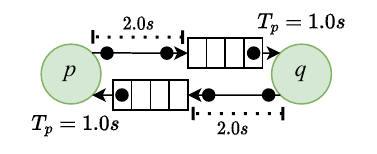}
        \caption{Tokens are in-flight and in the buffer, machines are live}
        \label[figure]{fig:delay_3_b}
    \end{subfigure}
    \caption{The example from \Cref{fig:delay_1}, with the initial marking changed to 3 tokens}
    \label[figure]{fig:delay_3}
\end{figure}

\highlight{Now consider that the number of initial tokens in each buffer is increased to three, shown in \Cref{fig:delay_3}. There are now sufficient tokens in circulation such that tokens are available to be consumed from each buffer at any time during execution. Consequently the transmission delay no longer causes unnecessary blocking, and both machines can operate at their nominal rate of 1.0 second.}

\highlight{
The optimal number of tokens (initial marking) is trivial to find in this example due to its symmetry. However, in the general case the differing execution rates of machines will always mandate some blocking. In these cases, the smallest marking resulting in maximum network performance is a special
case of a maximum-flow circulation problem~\cite{ahuja1988network}, which is NP-hard.}

\highlight{Introducing additional tokens is not behaviour-preserving for the application which is executing on each machine. In this case, the token-pushing systems are being used to model network-level synchronisation, where the contents of each token is indistinguishable (much like a petri net). The application model operates at a higher level of abstraction which will be covered in future works. The relationship between initial marking and throughput is well-known and not novel to this work. However, this pipelining property forms the basis of how the bittide model operates.}

\subsection{The Bittide Approach}

Recently, Google has developed a physical layer protocol called
bittide~\cite{lall_modeling_2021}.
Like \acp{FFP} bittide also uses point to point \ac{FIFO} buffers
(called \emph{elastic buffers}) for communication.

bittide makes use of token pipelining in the fashion described above, making it both performant and deterministic. However, in bittide, the computation of a process (its execution model) is decoupled from its synchronisation model with other processes, such that blocking should never occur.
Each process executes using its local clock.
When the clock ticks, the sender process
writes to its elastic buffer.
Likewise, when the consumer clock ticks, a token is removed
from its \ac{FIFO} buffer.

It is obvious that mismatching frequencies of the
producer and consumer will cause buffer overflow / underflow. However, in bittide,
this is handled by distributed controllers at each process. The controllers examine
the local buffer at a process to determine the relative speed of the neighbours.
Using this as a feedback signal, appropriate controllers are designed, which
ensure that frequencies of all processes stabilise to a common frequency,
ensuring buffer boundedness.

This protocol has been shown to be both deterministic and performant. However, bittide is still very recent and hasn't been thoroughly examined \highlight{either formally or experimentally. This paper aims to provide a formal model for bittide, and to compare it with existing models of computation.}

\section{Related Work}
\label{sec:related}
\highlight{The most common approach to distributed process synchronisation is through physical time protocols. Notably, the Network Time Protocol (NTP)~\cite{mills1991internet} synchronises physical clocks across a network to a common time source. More precise protocols in this category include the Precision Time Protocol (PTP)~\cite{ptp} and the recent IEEE 802.1AS~\cite{tsn} standard, which requires hardware compatibility. However, none of these actually describe how processes should synchronise, only how they should agree on a common time.} 
\highlight{The recent DetNet~\cite{detnet} standard builds over these precision protocols to ensure the deterministic delivery of data over a network, but still at a much lower level than process synchronisation.}

\highlight{The family of \acp{TTA}~\cite{kopetz2003time} provide a mechanism for process synchronisation based on physical time. These are commonly used for safety-critical industrial systems and provide both clock synchronisation and a mechanism for process synchronisation. However, their initial design complexity is quite high, and relies on a well-defined hardware structure such that they are not easily adapted for general-purpose distributed systems.}

At a higher level of abstraction, the coordination language Lingua Franca~\cite{linguafranca} enables both physical time and process synchronisation over arbitrary hardware. Its physical synchronisation protocol is based on the earlier Ptides~\cite{derler2008ptides}.
Using known latency bounds, each process can locally decide a ``safe-to-process''
time to proceed. When these bounds are violated, the system flags a
fault and invokes a handler.
While robust, Lingua Franca primarily focuses on the
mixture of logical and physical time to ensure the safe execution
of safety-critical system, rather than our
logical synchrony approach of abstracting away physical time entirely.

\highlight{Lamport clocks, and the related vector clocks, are the most famous logical time protocols~\cite{lamport2019time}. These logical clocks ensure that the order of events is consistently agreed on by all processes. Programmers may implement algorithms using these clocks to ensure process synchronisation, but that is still ultimately up to the designer to implement.}

\Acp{LTTA}~\cite{baudart_loosely_2016} describe a family of protocols, which preserve synchronous semantics over quasi-periodic architectures. \ac{LTTA} protocols are typically described at a much higher level than their namesake \acp{TTA}, focusing more on process-level communication. While the specific protocol varies, all designs synchronise logical tick progression through
the use of explicit messaging.
As a result, throughput degrades with higher transmission delays. In general, the communication delays are assumed to be small relative to the task duration.

\Ac{GALS} models are those which preserve synchronous semantics within a single machine, but communicate over an asynchronous medium. \highlight{The term \Ac{GALS} itself is very broad, often describing any system which combines synchronous and asynchronous components, but also often specific programming paradigms\cite{malik2010systemj}}. Usually such designs introduce non-determinism at a system level. Determinism is often enforced over a GALS architecture through the synthesis of wrappers which stall a process when an unresolved dependency is reached. \highlight{This is commonly employed in the Polychrony~\cite{gautier2023polychronous} toolkit of the SIGNAL synchronous language~\cite{benveniste1991synchronous}. A similar approach was employed by the Esterel tool `ocrep'~\cite{caspi1999automatic}.} These processes which remain behaviourally correct whether distributed over a synchronous or an asynchronous medium are designated \textit{endochronous}. Determining endochrony in general is a difficult task, \highlight{and if frequent synchronisations are required then the system faces the same performance pitfall as \acp{LTTA} when lengthy transmission delays are present.}

\highlight{In short, various approaches to process synchronisation exist within a variety of contexts. However, these are rarely general-purpose and often require significant designer input either at the hardware or software level.}

We make the following contributions for the design of
deterministic and performant distributed systems:

\begin{enumerate}
    \item We formalise \acp{LSN}, recently introduced in~\cite{lall_logicalsync},
          in the context of deterministic \acp{MoC} for distributed systems.
    \item We show \acp{KPN} are \iac{LSN} instance.
    \item  We show that
          bittide~\cite{lall_modeling_2021} is also a realisable \ac{LSN} instance.
          We introduce a variant of \acp{FFP} called \acp{LSFP} for making a fair comparison
          with bittide (see \Cref{sec:masking}).
    \item We present an empirical evaluation of the performance of
          \acp{LSFP} and bittide, and we show that bittide is more performant
          in the average case.
\end{enumerate}

\noindent This paper is organized as follows: in \Cref{sec:logicalsynchrony}, we introduce our definition of \acp{LSN}, adapted from~\cite{lall_logicalsync}.
Two realisable implementations are demonstrated.
In \Cref{sec:kpn} we show that the well-known \ac{KPN}
model can be used to express \acp{LSN}. In \Cref{sec:FFP} we show how \acp{FFP} implement \acp{KPN}, and thus propose \acp{LSFP} as a restriction which implements the \ac{LSN} model efficiently. Similarly, the recent physical-layer protocol bittide is
shown to express \acp{LSN} in \Cref{sec:bittide}.
Then, \Cref{sec:analysis} compares the performance of \acp{FFP} relative to bittide.
Finally, \Cref{sec:conclusion} makes concluding remarks including future directions.

\section{Logical Synchrony Networks}
\label{sec:logicalsynchrony}
\ignore{
    \Acp{LSN}, introduced in~\cite{lall_logicalsync}, describe a
    model for the deterministic execution of distributed systems
    by leveraging invariant (i.e. constant) logical delays, between every producer-consumer pair.
    We formalise the \ac{LSN} model, in this paper,
    to show that existing models of computations, especially those based on \acp{KPN},
    exhibit the \ac{LSN} property.
}

%\subsection{Preliminaries}
We provide a generic, graph-based abstraction of a distributed
system comprising a network of machines $\mathscr{M}_i$,
denoting the $i-th$ machine, which
labels the vertex $v_i$.
Furthermore, we formalise the \ac{LSN} model in this paper, \highlightg{which was introduced briefly in~\cite{lall_logicalsync} but not formalised as a \ac{MoC}},
and we show that existing models of computations exhibit the \ac{LSN} property of \emph{invariant logical delays}, especially those based on \acp{KPN}.

Machines execute at discrete points called events or ticks.
Each machine has its own notion of ticking, which may be different from other machines.
The event count $\theta_i\in \mathbb{N}$, corresponding to the
machine $\mathscr{M}_i$ increases by 1 every logical tick.
The defining feature of \acp{LSN} is the
\textbf{invariant logical delays}, meaning there always exists
an invariant offset between the event count of any production
events and their associated consumption events along a channel.

An event is denoted ($\mathscr{M}_i$, $\theta_i$), where $\mathscr{M}_i$
is the machine experiencing the event when $\theta_i$ is its event counter.
We say that two events $ (\mathscr{M}_i, \theta_i) \prec (\mathscr{M}_j, \theta_j)$
iff $\theta_j - \theta_i = \lambda_{i\rightarrow j}$.
At each event, a token is read from each input edge
and a token is produced to each output edge.
The event count $\theta_i\in \mathbb{N}$ increases by 1.

\begin{definition}
    \Iac{LSN} is a tuple $<G, \Theta, \lambda>$, where:

    \begin{itemize}[]
        \item $G{=}{<}V,E{>}$ is a digraph  where $V$ denotes a set of vertices,
              and $E$ $\subseteq V {\times} V$ denotes a set of edges
              such that $v1 \neq v2$ for all $(v1, v2) \in E$.
        \item Each vertex $\mathscr{M}_i \in V$ corresponds to a
              machine executing a synchronous program that generates
              an event every logical tick.
              $\theta_i \in \Theta$ represents the event counter for the machine
              $\mathscr{M}_i$ and
              the value of $\theta_i \in \mathbb{N}$.
              As the next event is generated by a machine as its \emph{ticks},
              its event counter is incremented by 1.
        \item Edges are labelled with the logical delay
              mapping  $\lambda : E \rightarrow \mathbb{Z}$. Moreover,
              $\lambda((\mathscr{M}_i,\mathscr{M}_j)) = \lambda_{i\rightarrow j}$ implies
              the following relation holds between the events associated with this edge:
              $ (\mathscr{M}_i, \theta_i) \prec (\mathscr{M}_j, \theta_j)$.
              Consequently, $\theta_j - \theta_i = \lambda_{i\rightarrow j}$.
    \end{itemize}
\end{definition}

\begin{example}
    An example \ac{LSN} is shown in \Cref{fig:lsn_example}, consisting of three machines
    $M_i, M_j, M_k$. An example edge is between $(M_k, M_j)$, which is labelled with
    a value 2, which indicates the invariant logical delay, denoted $\lambda_{k\rightarrow j}$,
    between production of
    some event at $M_k$ and consumption at $M_j$ at any tick of the two machines when the
    production and the corresponding consumption happens. For this edge $\lambda_{k\rightarrow j} =2$.

    \begin{figure}[!htbp]
        \centering
        \includegraphics[clip, trim=0.1cm 0.1cm 0.1cm 0.0cm,width=0.6\linewidth]{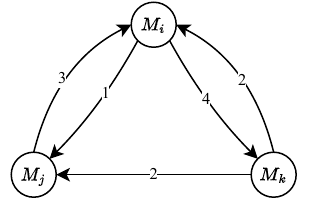}
        \caption{A three-node \acs*{LSN}}
        \label[figure]{fig:lsn_example}
    \end{figure}
\end{example}

The defining feature of \acp{LSN} is the
\emph{invariant logical delays}, meaning there always exists
an invariant offset between the event count of any production
events and their associated consumption events along a channel. The same invariance applies over both edges and paths.
Multiple paths, with different cumulative logical delays, may exist between any two machines. $\lambda_{i\rightarrow{j}}$.
This does \textbf{not} imply that the machines have
any relationship between event counts at any one physical time
when observed by a global entity; any amount of physical time may have elapsed between the two events. \highlight{Logical delays are defined over the set of integers, including negative values. Whether this is used in a physical system is an implementation detail. Lall et. al ~\cite{lall_logicalsync} define an equivalence class of \acp{LSN}, and in doing so show that all \acp{LSN} with negative edge weights have at least one equivalent all-positive \ac{LSN}.}

\begin{example}
    \Cref{fig:example_timeline} shows an event
    sequence for the \ac{LSN} in \Cref{fig:lsn_example} with a potential execution trace
    of the three machines and their relationship in both logical and physical time.
    The three machines are desynchronised in their event timings
    in physical time. However, the logical time offset remains invariant.

    \begin{figure}[H]
        \centering
        \includegraphics[clip, trim=0.5cm 0.3cm 0cm 0.3cm,width=0.9\linewidth]{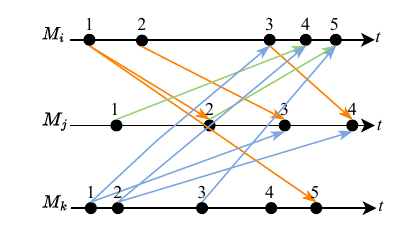}
        \caption{The execution trace of the three-machine \acs*{LSN} (\Cref{fig:lsn_example}), as measured by a global observer}
        \label{fig:example_timeline}
    \end{figure}
\end{example}

\highlight{Here we show that \acp{LSN}, as a model, are deterministic:} In order to reason about \ac{LSN} behaviours, we start by abstracting machines as mathematical functions.
Each machine $\mathscr{M}_i\in V$ has a corresponding function
$f_i \in \mathscr{F}$, where $\mathscr{F}$ is the set of all machine functions,
which can be described as a synchronous Mealy machine:

\begin{equation}
    f_i : X^n \times S_i \rightarrow X^m \times S_i
\end{equation}

Where $X^n$ denotes a vector of $n$ inputs and $X^m$ denotes a vector of $m$ output values from that machine.
$S_i$ denotes the set of states for $\mathscr{M}_i$.
The function takes in all current inputs and the current state, and produces some outputs and the next state.

\begin{figure}[htp]
    \centering
    \includegraphics[clip, trim=0.6cm 0.05cm 0.6cm 0.22cm,width=0.55\linewidth]{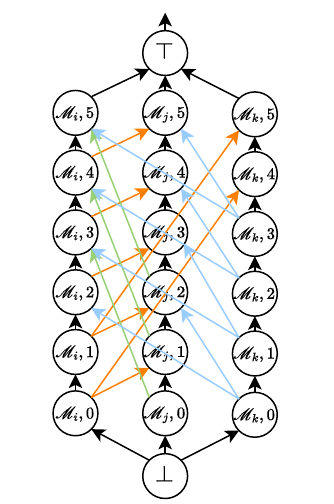}
    \caption{$G_{ext}$ of the three-machine \acs*{LSN} (\Cref{fig:lsn_example}). Computation edges shown in black and communication edges coloured to differentiate source machine}
    \label{fig:extendedgraph}
\end{figure}

\begin{definition}
    \Iac{LSN} execution forms \iac{CPO} of events, called
    the extended graph $G_{ext}={<}V_{ext},E_{ext}{>}$.
    \begin{itemize}
        \item $V_{ext}$ is the set of all machine events in the network, described by $V_{ext} = \{(\mathscr{M}_i,n) | \mathscr{M}_i \in V, n \in \theta_i\}$
        \item $E_{ext}$ describes the directed edges showing dependencies between events, which are of one of two types.
              \begin{enumerate}
                  \item Edges between events at different machines $(\mathscr{M}_i,n) {\rightarrow}(\mathscr{M}_j,m)$ are called communication edges, corresponding to an edge $(\mathscr{M}_i,\mathscr{M}_j) \in E$ in the \ac{LSN}. The clock difference $m{-}n$ is the associated logical delay $\lambda_{i\rightarrow j}$
                  \item Edges between successive events at the same machine $(\mathscr{M}_i,n) {\rightarrow}(\mathscr{M}_i,n+1)$ are called computation edges, which capture monotonicity of local machine events.
              \end{enumerate}
    \end{itemize}
    \label{lemma:cpo}
\end{definition}
Due to the condition on \acp{LSN} that all cycles in the graph are positive, the corresponding ordering relation is monotonic and acyclic by construction. $G_{ext}$ for the example shown in \Cref{fig:lsn_example} is demonstrated in \Cref{fig:extendedgraph}. We assume a finite execution, meaning there is a unique start-up event $\bot$, which precedes the first event at each machine, and termination event $\top$, which succeeds the final event at each machine.

Each event is associated with a corresponding function instantiation,
where $f_i^0$ represents the first invocation of machine $\mathscr{M}_i$
at event $(\mathscr{M}_i,0)$, and $f_i^{k-1}$ the $kth$
invocation at $(\mathscr{M}_i,k-1)$.
The input and output edges of an event form
the input and output vectors respectively and the state corresponds to the
event. Naturally, an event cannot receive logically delayed input from a non-existing negative event, so to accommodate missing inputs we introduce initial conditions within $\bot$ containing the pre-calculated inputs to the initial function invocations.

Determinism has a variety of different interpretations in literature
as summarised in~\cite{edwards2018determinism}. In this context we interpret it as follows:
\begin{definition}
    \Iac{LSN} system is determinate if for any two or more executions with consistent initial conditions, the resulting output traces are consistent, where:
    \begin{itemize}
        \item The output derived from the initial condition consists of the function instance applied to the initial inputs $\bot$.
        \item An output trace for a single machine consists of the sequence of outputs from each function instance $\langle \mathbf{f^i} \rangle = \langle f_i^0::f_i^1::...::f_i^{n-1} \rangle $ where $::$ is the concatenation operator.
    \end{itemize}
\end{definition}

\begin{lemma}
    Consider any event $(M_i,n) \in V_{ext}$.
    The value of any edge (incoming/outgoing) is determinate.
\end{lemma}
\begin{IEEEproof}
    Any edge is a composition of functions and hence determinate.
\end{IEEEproof}

\begin{theorem}
    \ac{LSN} execution is always determinate.
\end{theorem}

\begin{IEEEproof}
    Based on Lemma 1 and by induction on the depth $k$ of the extended
    graph corresponding to the \ac{LSN}. Note that any event is only dependent on
    events that happened prior to it.
\end{IEEEproof}

\highlight{Of course, such guarantees are only valid for the abstract model of \acp{LSN}. In practice, the implementation of the machines must be proven to conform to the abstract model. Specifically, we show this for the \ac{KPN} and bittide implementations in the following sections.}

\section{KPN as an instance of LSN}
\label{sec:kpn}
\Acp{KPN} are a well-known model of computation for deterministic systems. \highlight{We show that this existing \ac{MoC} satisfies our \ac{LSN} model.} \Acp{KPN} model a network of machines communicating over unbounded \ac{FIFO} channels.  Each machine in \iac{KPN} executes over discrete firings, analogous to \ac{LSN} events, where at each firing a token is consumed from every inbound \ac{FIFO}, and a token pushed onto each outbound \ac{FIFO}.
\begin{definition}
    \Iac{KPN} is denoted as a graph $G'=<V',E'>$ and a labelling function $\alpha :  E' \rightarrow B$ where any $b \in B$ is an unbounded FIFO queue. Each buffer has an associated occupancy of tokens given by $\beta : B \times \mathbb{R} \rightarrow \mathbb{N}$, where $\mathbb{R}$ is the current physical time and $\mathbb{N}$ the count of tokens in the \ac{FIFO}.
\end{definition}

\begin{figure}[htp]
    \centering
    \includegraphics[clip, trim=0cm 0.02cm 0cm 0.02cm,width=0.60\linewidth]{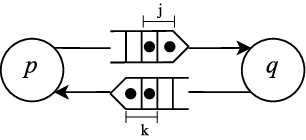}
    \caption{A cyclical \acs*{KPN}, with some initial tokens in each directional unbounded \acs*{FIFO}}
    \label{fig:loopkpn}
\end{figure}

A basic two-vertex \ac{KPN} is shown in \Cref{fig:loopkpn}, with the initial occupancies $\beta(\alpha(p,q),0)=j$ and $\beta(\alpha(q,p),0)=k$. \Iac{KPN} with cyclic components will immediately deadlock if there are no initial tokens. Every directed cycle in a graph must always have at least one token to be live. We call these values the \textit{initial marking}.

\Acp{KPN} alone do not describe when a machine may fire, so we introduce a set of \textit{firing rules} which dictate firing times. Numerous firing orders may be possible with different temporal interleaving of events, as long as \acp{FIFO} do not underflow. Conventionally this is avoided through use of blocking channel reads.

By observation of the graphical similarities between \acp{LSN} and \acp{KPN}, and the effect of initial \ac{FIFO} markings on delays, we can deduce a simple transformation:
\begin{observation}
    \label{obs:transformkpn}
    Given \iac{KPN} we can produce an equivalent \ac{LSN} using the following algorithm:
\end{observation}
\begin{enumerate}
    \item For the \ac{KPN} graph $G'{=}{<}V',E'{>}$, form \iac{LSN} graph $G{=}{<}V,E{>}$. For every $v' \in V'$, there exists a $v \in V$, and for each $e' \in E'$, there exists an $e \in E$.
    \item Label each edge $(p,q) \in E$ with the logical delay $\lambda_{p\rightarrow q}$ = $\beta_{p\rightarrow{}q}(0)$, equal to the initial occupancy of the \ac{FIFO} in the original \ac{KPN}.
\end{enumerate}

Consider the \ac{FIFO} in \Cref{fig:loopkpn}. There exist $j$ initial tokens from $p\rightarrow{}q$. As a result, the first token produced at $p$ will be consumed by $q$ on its $(j+1)_{th}$ firing, as it must first await all initial tokens to be consumed.
\begin{observation}
    The logical delay $\lambda_{p\rightarrow{}q}(t)$ at any time $t$ consists of the \ac{FIFO} occupancy $\beta_{p\rightarrow{}q}(t)$, plus the fire count difference $\theta_{q}(t){-}\theta_{p}(t)$.
    \begin{equation*}
        \begin{gathered}
            \lambda_{p\rightarrow{}q}(t)=\beta_{p\rightarrow{}q}(t)+\theta_{q}(t)-\theta_{p}(t)
        \end{gathered}
    \end{equation*}
\end{observation}

\begin{lemma}
    \label{lemma:kpninvar}
    The logical delay $\lambda_{p\rightarrow{}q}(t)$ is invariant for all physical wall-clock times $t$:
    \begin{IEEEproof}

        \noindent\textit{\underline{Case 1:} } t=0:
        \begin{IEEEeqnarray}{c}
            \lambda_{p\rightarrow{}q}(0)=\beta_{p\rightarrow{}q}(0)+\theta_{q}(0)-\theta_{p}(0) \IEEEnonumber\\
            \lambda_{p\rightarrow{}q}(0)=\beta_{p\rightarrow{}q}(0)+0-0 = \beta_{p\rightarrow{}q}(0)\IEEEnonumber
        \end{IEEEeqnarray}
        \textit{\underline{Case 2:} p fires after time $\tau_p$, producing a token}
        \begin{IEEEeqnarray*}{c}
            \beta_{p\rightarrow{}q}(t+\tau_p) = \beta_{p\rightarrow{}q}(t)+1 \IEEEnonumber\\
            \text{The firing count of $p$ increases:}\IEEEnonumber\\
            \theta_{p}(t+\tau_p) = \theta_p(t) + 1 \IEEEnonumber\\
            {\lambda_{p\rightarrow{}q}(t+\tau_p)=(\beta_{p\rightarrow{}q}(t){+}1)+(\theta_{q}(t))-(\theta_p(t){+}1)} \IEEEnonumber\\
            \lambda_{p\rightarrow{}q}(t+\tau_p) = \beta_{p\rightarrow{}q}(t)+\theta_{q}(t)-\theta_p(t) \IEEEnonumber\\
            \therefore \lambda_{p\rightarrow{}q}(t+\tau_p) = \lambda_{p\rightarrow{}q}(t)\IEEEnonumber
        \end{IEEEeqnarray*}
        \textit{\underline{Case 3:} q fires after time $\tau_q$, consuming a token}
        \begin{IEEEeqnarray*}{c}
            \beta_{p\rightarrow{}q}(t+\tau_q) = \beta_{p\rightarrow{}q}(t)-1 \\
            \text{The firing count of $q$ increases:}\\
            \theta_{q}(t+\tau_q) = \theta_q(t)+1 \\
            \lambda_{p\rightarrow{}q}(t+\tau_q) = (\beta_{p\rightarrow{}q}(t){-}1) + (\theta_{q}(t) {+} 1) - (\theta_p(t)) \\
            \lambda_{p\rightarrow{}q}(t+\tau_q)  =\beta_{p\rightarrow{}q}(t) + \theta_{q}(t)- \theta_p(t) \\
            \therefore \lambda_{p\rightarrow{}q}(t+\tau_q) = \lambda_{p\rightarrow{}q}(t)
        \end{IEEEeqnarray*}
        \textit{\underline{Case 4:} p and q both fire simultaneously}
        \begin{IEEEeqnarray*}{c}
            \text{Linear combination of 2 and 3, invariance holds}. \qquad\IEEEQEDhere
        \end{IEEEeqnarray*}
    \end{IEEEproof}
\end{lemma}

The value of the logical delay remains invariant; the composition of this invariant value is exchanged between the buffer occupancy and the relative difference of fire counts. Henceforth we drop the time variable from the invariant  $\lambda_{pq}$.

\begin{theorem}
    Every \ac{KPN} is \iac{LSN}.
\end{theorem}
\begin{IEEEproof}
Follows from \Cref{obs:transformkpn} and \Cref{lemma:kpninvar}
\end{IEEEproof}
% \begin{theorem}
%     Not every \ac{LSN} is \iac{KPN}.
% \end{theorem}
% \begin{IEEEproof}\acp{LSN} admit negative logical delays, but \iac{FIFO} cannot have a negative marking.\end{IEEEproof}

There are difficulties mapping \iac{KPN} model to a physical implementation due to the unbounded \acp{FIFO}. \Ac{SDF}~\cite{lee_dataflow_1995} overcomes this limitation by generating a schedule of predetermined production/consumption events, but implementing efficient scheduling on distributed devices is challenging. Here we present two concrete implementations of \acp{LSN}, each with a different mechanism to bound buffer sizes.

\section{\highlightg{Finite FIFO Platforms}}
\label{sec:FFP}

\begin{figure}[htp]
    \centering
    \includegraphics[width=0.75\linewidth]{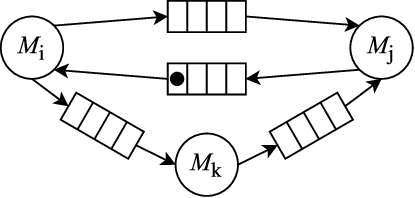}
    \caption{A three-machine \acs*{FFP} where one token is placed in the unit delay buffer to ensure liveness}
    \label{fig:ffp_diagram}
\end{figure}

\acp{FFP} are a realisation of the \ac{KPN} model, extended with blocking writes to prevent unbounded growth, \highlightg{first introduced in \cite{tripakis_implementing_2008}.} \highlight{We adopt them here as an example of an implementable \ac{LSN} model from existing literature.} 
A process can only fire if tokens are available for consumption at every input, and if there is space to emit tokens at every output.  Otherwise it `stutters', and skips that firing cycle. \Cref{fig:ffp_diagram} shows \iac{FFP} with a single initial token.

As a skipped cycle does not affect the semantics of \iac{KPN}, it is acceptable to stutter at any time. Just as in the pure \ac{KPN} model, $\lambda_{p\rightarrow{q}}$ is determined by the number of initial tokens from $p$ to $q$ to three.

In this approach, buffer boundedness is enforced by locating the intermediate buffer on the receiver side of any communication channel. Blocking reads are trivial to implement by checking the local buffer occupancy. Blocking writes are enforced by using a heuristic to conservatively estimate remote buffer occupancy. \highlight{This heuristic is mentioned in \cite{tripakis_implementing_2008} but explained in greater detail here.}

\subsection{Efficient Fullness Checking}
Consider two machines $p$ and $q$, separated by a transmission link with latency $l_{pq}$. The real occupancy of some buffer $\beta_{real}(t)$ at time $t$ is equal to the number of tokens that have arrived over the link from a producer $p$, minus the number consumed at $q$, plus the initial occupancy:
\begin{equation}
    \beta_{real}(t) = \theta_p(t-l_{pq}) - \theta_q(t) + \beta(0)
    \label{true_beta}
\end{equation}

\begin{figure}[htp]
    \centering
    \includegraphics[width=0.65\linewidth]{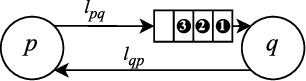}
    \caption{A buffered communication link from producer $p$ to consumer $q$, with a back-pressure link from $q$ to $p$}
    \label{fig:backreadlink}
\end{figure}

The producer $p$ cannot know the current value of $\theta_q(t)$ nor the number of its own tokens that have reached the remote buffer $\theta_p(t-l_{pq})$. To provide this information, we include an unbuffered reverse link from consumer to producer. \Cref{fig:backreadlink} demonstrates this arrangement.

Over this reverse link $p$ measures the delayed firing count $\theta_q(t-l_{qp})$. $p$ can also measure its own current firing count $\theta_p(t)$. The difference between these two counts is an estimate of the worst-case number of tokens that may still be in transit between the two machines, thus by substituting both of these into \Cref{true_beta}:

\begin{equation}
    \begin{gathered}
        \theta_p(t) \geq \theta_p(t-l_{pq})\\ \text{and } \theta_q(t-l_{qp}) \leq \theta_q(t)\\
        \therefore \beta_{true}(t) \leq \theta_p(t) - \theta_q(t-l_{qp}) + \beta(0)
    \end{gathered}
    \label{eqn:bworst}
\end{equation}

\begin{figure}[htp]
    \centering
    \includegraphics[clip, trim=0.55cm 0.0cm 0cm 0cm,width=0.9\linewidth]{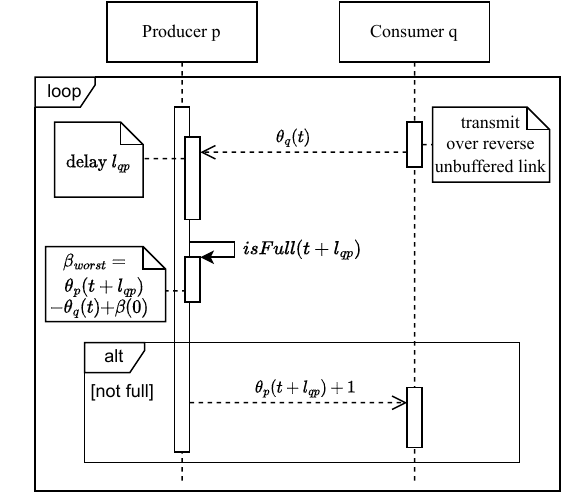}
    \caption{Sequence of isFull() \acs*{FFP} behaviour}
    \label{fig:sequence}
\end{figure}

Thus we can define a fullness checking function at $p$ which always gives a conservative occupancy estimate. This flow is given by the sequence diagram shown in \Cref{fig:sequence}.
The overestimation error is proportional to latency, so the peak buffer occupancy will be less than the full capacity.

\subsection{Modified FFPs for enhanced throughput}
\label{sec:masking}
One limitation of the \ac{FFP} model is that our performance degrades as we introduce communication delay, because the tokens required to progress spend some time in-flight. As discussed earlier, if a system has more initial tokens in circulation processes can execute more often due to pipelining. At some token count we reach a saturation the throughput does not increase with additional tokens because the communication link is fully populated with in-flight tokens. Such is the case in the bittide model, where the link is assumed to always be fully populated. As a result, a bittide system may have large logical delays ($\lambda$), but no blocking is required while awaiting inputs. Here we take inspiration from the delay masking of the bittide model and apply it to the \ac{FFP} model, and introduce a special class of \acp{FFP} which we denote as \acp{LSFP}:

\begin{definition}
    \Iac{LSFP} is a special case of \iac{FFP} where the initial marking in each buffer is governed by a heuristic as follows used to enhance throughput:

    \begin{itemize}[]
        \item For an edge $(\mathscr{M}_i,\mathscr{M}_j) \in E$
              with some transmission time $l_{i\rightarrow j}$ and
              a consumer frequency $\omega_j$, the initial occupancy $\beta_{i\rightarrow j}(0)=\frac{l_{i\rightarrow j}}{\omega_j}$
    \end{itemize}
\end{definition}

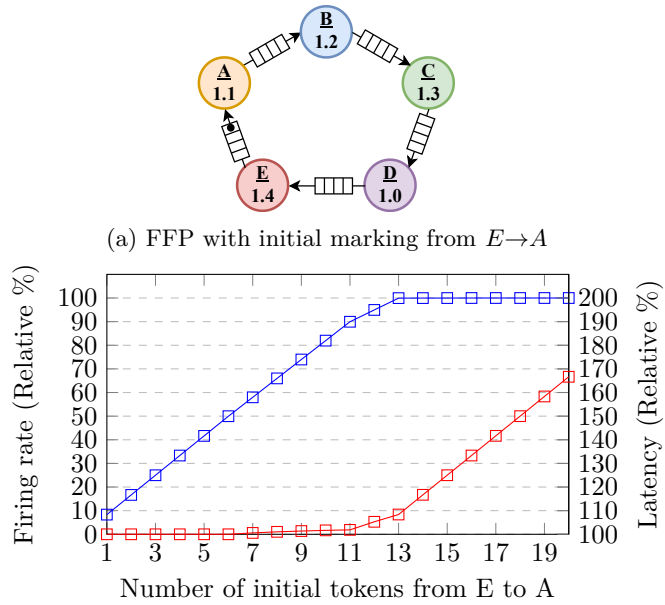
\begin{figure}[htp]
    \centering
    \begin{subfigure}[t]{\linewidth}
        \centering
        \includegraphics[clip, trim=0.0cm 0.1cm 0.0cm 0.1,width=0.4\linewidth]{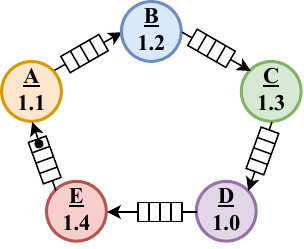}
        \caption{\acs*{FFP} with initial marking from $E{\rightarrow}A$}
        \label{fig:ring_example:graph}
    \end{subfigure}
    \begin{subfigure}[t]{\linewidth}
        \centering
        \begin{tikzpicture}
	\begin{axis}[
		xlabel={Number of initial tokens from E to A},
		ylabel={Firing rate (Relative \%)},
		xmin=1, xmax=20,
		ymin=0, ymax=110,
		xtick={1,3,5,7,9,11,13,15,17,19},
		ytick={0,10,20,30,40,50,60,70,80,90,100},
		legend style={inner sep=0pt},
		legend pos=north west,
		ylabel near ticks, yticklabel pos=left,
		ymajorgrids=true,
		grid style=dashed,
		width=0.9\linewidth,
		height=5cm
		]
		\addplot[
		color=blue,
		mark=square,
		]
		coordinates {
			(1,0.0833*100)
			(2,0.1666*100)
			(3,0.2499*100)
			(4,0.3333*100)
			(5,0.4165*100)
			(6,0.49977*100)
			(7,0.5797*100)
			(8,0.6598*100)
			(9,0.7397*100)
			(10,0.8196*100)
			(11,0.8995*100)
			(12,0.9495*100)
			(13,0.9995*100)
			(14,0.9997*100)
			(15,0.9998*100)
			(16,1*100)
			(17,1*100)
			(18,1*100)
			(19,1*100)
			(20,1*100)
		};
		
	\end{axis}
	\begin{axis}[
		axis y line*=right,
		axis x line=none,
		xmin=1, xmax=20,
		ymin=100, ymax=210,
		ylabel near ticks, yticklabel pos=right,
		legend style={inner sep=0pt},
		xtick={1,3,5,7,9,11,13,15,17,19},
		ytick={100,110,120,130,140,150,160,170,180,190,200},
		ylabel={Latency (Relative \%)},
		width=0.9\linewidth,
		height=5cm
		]
		\addplot[
		color=red,
		mark=square,
		]
		coordinates {
			(1,2.4*100/2.4)
			(2,2.4*100/2.4)
			(3,2.4*100/2.4)
			(4,2.4*100/2.4)
			(5,2.4*100/2.4)
			(6,2.4*100/2.4)
			(7,2.414*100/2.4)
			(8,2.425*100/2.4)
			(9,2.433*100/2.4)
			(10,2.440*100/2.4)
			(11,2.445*100/2.4)
			(12,2.527*100/2.4)
			(13,2.601*100/2.4)
			(14,2.801*100/2.4)
			(15,3.000*100/2.4)
			(16,3.200*100/2.4)
			(17,3.400*100/2.4)
			(18,3.600*100/2.4)
			(19,3.800*100/2.4)
			(20,4.000*100/2.4)
		};
	\end{axis}
\end{tikzpicture}
        \caption{Firing rate in blue (higher is better) and latency in red (lower is better) versus initial marking}
        \label{fig:ring_example:results}
    \end{subfigure}

    \caption{A ring \acs*{FFP} network with frequencies labelled on each node. Links have a delay of 2.0 seconds.}
    \label{fig:ring_example}
\end{figure}

The increase in throughput as token count increases is demonstrated for
illustrative purposes in \Cref{fig:ring_example} for a simulated \ac{FFP} (see \Cref{sec:analysis}).
This topology was selected arbitrarily, but the expected trend is not topology-specific.
Note that the physical response time is not meaningfully worsened until after the saturation point.
Now, each process becomes much closer to free-running execution. Consequently, predictable execution is gained with minimal impact to efficiency compared to an asynchronous system. These logical delays may be used in interesting ways at an application level, discussed in future works.

\section[bittide]{\MakeLowercase{\highlightg{bittide}}}
\label{sec:bittide}
bittide~\cite{lall_modeling_2021,lall_resistance_2021} is a recent physical-layer protocol for distributed computing, developed at Google Research. \highlight{Bittide serves as the inspiration for this more abstract \ac{LSN} \ac{MoC}.} Inspired by elastic circuits~\cite{carmona_elastic_2009}, variations in frequency are absorbed by an intermediate \ac{FIFO} to maintain syntony.

In a bittide network, a communication link corresponding to an
edge in \iac{LSN} between some $\mathscr{M}_i$ and $\mathscr{M}_j$ is implemented by the following:
\begin{enumerate}
    \item A physical communication link, holding a number of in-flight frames called link occupancy $\gamma_{i\rightarrow{}j}$
    \item \Iac{FIFO} co-located at the receiver with current occupancy $\beta_{i\rightarrow{}j}$
\end{enumerate}

Periodically, a clock at a machine $\mathscr{M}_i$ will `tick', increasing its logical clock count $\theta_i$ by one.
\begin{equation*}
    \theta_i = 	\{0,1,2\ldots\} \in \mathbb{N}
\end{equation*}

% In contrast to the firing model of \acp{KPN}, bittide clocks are not assumed to have a coordinated startup. 

At each tick, a machine consumes a frame (token) from each inbound \ac{FIFO}, and emits a frame on each outbound link. Each link will at any point hold some frames in flight, denoted by the link occupancy. When a frame of data arrives at $\mathscr{M}_j$ from the inbound link it appends to the tail of the recipient \ac{FIFO}. It will be later consumed by the receiver machine once it propagates to head of the queue after $\beta_{i\rightarrow{}j}$ receiver ticks.

Compared to \iac{KPN} approach, the main feature of bittide is that each machine remains completely free running, rather than blocking. This is possible due to a dynamic clock control which balances frequencies.
\subsection{Clock Control System}
\label{sec:clockcontrol}
\begin{figure}[htp]
    \centering
    \includegraphics[clip, trim= 0.0cm 0.0cm 0.0cm 0.18cm, width=0.95\linewidth]{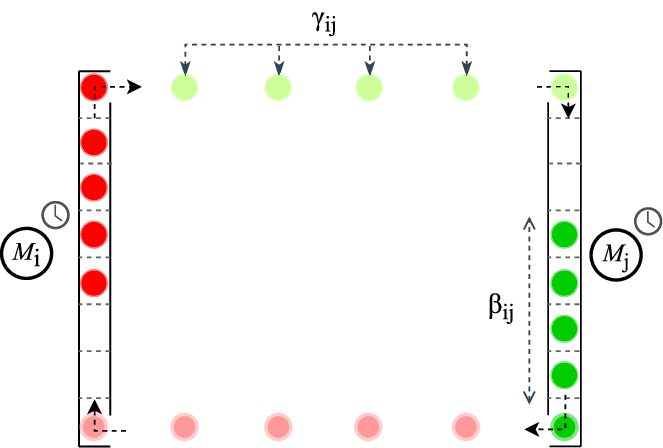}
    \caption{Interaction between two bittide machines, demonstrating the \acs*{FIFO} and link action}
    \label{fig:twonode}
\end{figure}

Typically, bittide implementations assume bidirectional links, each edge in the \ac{LSN} therefore forming a circuit of token flow. \Cref{fig:twonode} shows the communication loop between two machines $\mathscr{M}_i$ and $\mathscr{M}_j$ in a network.

The current occupancies of a machine's buffers are used as a feedback signal to determine clock rate relative to its neighbours, and thus a clock control policy can be applied to take corrective action and stabilise the buffer occupancy within reasonable bounds. The control system is designed with the goal of preventing buffer overflow or underflow, but does not impact the logical synchrony property (so long as frames aren't lost). Various control policies may be valid for a given network topology, such as using a proportional controller~\cite{lall_modeling_2021}, a proportional-integral controller~\cite{lall_resistance_2021}, and a novel ``reframing'' controller~\cite{lall2023buffer}. Bidirectional links are not a hard requirement for clock control, but have so far been assumed in previous works.

The free-running nature of bittide is that frames are assumed to always be in circulation. Unlike in the \ac{KPN} model with blocking reads, a bittide system should not stall while awaiting a dependent signal. There must be sufficient initial frames, and thus sufficiently deep logical delays, such that the elastic buffer is never completely exhausted while awaiting inbound frames.

\subsection{\highlight{Logical Delay Invariance}}
\highlight{We show that the bittide architecture satisfies our definition of \acp{LSN}.}
\begin{observation}
    A frame emitted from $\mathscr{M}_i$ at time \textit{t} will arrive at $\mathscr{M}_j$ at time  \textit{t+$\tau_{i\rightarrow j}(t)$} after all preceding frames on the path.
\end{observation}

$\tau_{i\rightarrow j}(t)$ represents the physical time elapsed up to consumption, consisting of link delay and the time spent queuing through the \ac{FIFO}. One frame is consumed per logical tick, therefore the clock $\theta_j$ will also have increased by the number of frames at $t$ on the $i\rightarrow j$ path, consisting of the link and the buffer:
\begin{equation}
    \begin{gathered}
        \theta_j(t+\tau_{i\rightarrow j}(t)) = \theta_j(t) + \gamma_{i\rightarrow{}j}(t) + \beta_{i\rightarrow{}j}(t)
    \end{gathered}
    \label{eqn:clockrelationshipj}
\end{equation}
\Cref{eqn:clockrelationshipj} describes the evolution of the consumer clock during one transmission. Let's re-write this to include an expression for the producer clock $\theta_i$:
\begin{equation}
    \begin{gathered}
        \text{let } \Delta\theta_{i\rightarrow j}(t) = \theta_i(t) - \theta_j(t)\\
        \text{then from \Cref{eqn:clockrelationshipj}:} \\
        \theta_j(t+\tau_{i\rightarrow j}) = \theta_i(t) + \Delta\theta_{i\rightarrow j}(t) + \gamma_{i\rightarrow{}j}(t) + \beta_{i\rightarrow{}j}(t)
    \end{gathered}
\end{equation}
Collect all terms except for the producer and consumer clock terms to define a relationship between them:
\begin{IEEEeqnarray}{c}
    \label[equation]{eqn:phaserelationship}
    {\theta_j(t+\tau_{i\rightarrow j}) = \theta_i(t) + \underbrace{\gamma_{i\rightarrow{}j}(t) + \beta_{i\rightarrow{}j}(t) + \Delta\theta_{i\rightarrow j}(t)}_{\textstyle\lambda_{i\rightarrow{}j}(t)\mathstrut}}\IEEEnonumber\\
    {\theta_j(t+\tau_{i\rightarrow j}) = \theta_i(t) + \lambda_{i\rightarrow{}j}(t)}
\end{IEEEeqnarray}

Thus, we define a logical delay $\lambda_{i\rightarrow{}j}(t)$, describing the offset between the clock at $\mathscr{M}_i$ when a frame is produced and the clock at $\mathscr{M}_j$ when consumed. To satisfy the \ac{LSN} definition, we show that $\lambda_{i\rightarrow{}j}(t)$ is invariant. $\lambda_{i\rightarrow{}j}(t)$ is invariant if the difference equation $\lambda_{i\rightarrow{}j}'(t)=(\lambda_{i\rightarrow{}j}(t+\tau_{i\rightarrow j})-\lambda_{i\rightarrow{}j}(t))$ is 0 for all send-receive physical time pairings:
\begin{IEEEeqnarray}{c}
    \text{from (\ref{eqn:phaserelationship}): } \lambda_{i\rightarrow{}j}'(t) = \gamma_{i\rightarrow{}j}'(t){+}\beta_{i\rightarrow{}j}'(t) {+} \Delta\theta'_{i\rightarrow j}(t)\IEEEnonumber\\
    \text{prove }\gamma_{i\rightarrow{}j}'(t){+}\beta_{i\rightarrow{}j}'(t) {+} \theta'_j(t){-}\theta'_i(t)=0
    \label{eqn:invariance}
\end{IEEEeqnarray}

\begin{observation}
    Line occupancy is incremented during a producer tick, and is decremented when arriving at recipient \ac{FIFO}.
\end{observation}
The line occupancy $\gamma_{i\rightarrow{}j}$ is equal to the number of frames that have been pushed onto the line by $\mathscr{M}_i$, minus the number that have arrived at the buffer, plus the initial condition. We separate these two terms as the cumulative number pushed $\gamma_{i\rightarrow{}j}^{in}$ and the cumulative number received $\gamma_{i\rightarrow{}j}^{out}$
\begin{IEEEeqnarray}{rCl}
    \text{let }\gamma_{i\rightarrow{}j}(t) = \gamma_{i\rightarrow{}j}^{in}(t) - \gamma_{i\rightarrow{}j}^{out}(t) + \gamma_{i\rightarrow{}j}(0)\IEEEnonumber\\
    \text{then difference } \gamma_{i\rightarrow{}j}'(t) = {\gamma_{i\rightarrow{}j}^{\prime\hspace{3pt}in}}(t) - \gamma_{i\rightarrow{}j}^{\prime\hspace{3pt}out}(t)
    \label{eqn:splitgamma}
\end{IEEEeqnarray}
As one frame being added to the link corresponds to one logical tick at $\mathscr{M}_i$, we can say that:
\begin{IEEEeqnarray}{rCl}
    \label[equation]{eqn:yp}
    {\gamma_{i\rightarrow{}j}^{\prime\hspace{3pt}in}}(t) = \theta_i'(t)\IEEEnonumber\\
    \text{from (\ref{eqn:splitgamma}) } \gamma_{i\rightarrow{}j}'(t) =  \theta_i'(t) - {\gamma^{\prime\hspace{3pt}out}_{i\rightarrow{}j}}(t)\IEEEnonumber\\
    \text{i.e. }\theta_i'(t) = {\gamma_{i\rightarrow{}j}'(t) + \gamma^{\prime\hspace{3pt}out}_{i\rightarrow{}j}}(t)
    \label[equation]{eqn:pbx}
\end{IEEEeqnarray}
\begin{observation}
    Buffer occupancy $\beta_{i\rightarrow{}j}$ decreases when $\theta_j$ ticks, and increases when a frame pops off the link.
    \begin{IEEEeqnarray}{rCl}
        \beta_{i\rightarrow{}j}'(t) =  {{\gamma_{i\rightarrow{}j}^{\prime\hspace{3pt}out}}(t)-\theta_j'}(t)\IEEEnonumber\\
        \text{i.e. }\theta_j'(t) = {\gamma_{i\rightarrow{}j}^{\prime\hspace{3pt}out}}(t) -\beta_{i\rightarrow{}j}'(t)
        \label[equation]{eqn:pby}
    \end{IEEEeqnarray}
\end{observation}

We now have expressions for $\theta_i'(t)$ and $\theta_j'(t)$.

\begin{lemma}
    Logical delay $\lambda_{i\rightarrow{}j}(t)$ is invariant for all $t$:
    \label{lemma:1}
    \begin{IEEEproof}
        \begin{IEEEeqnarray}{rCl}
            \text{\normalfont{}Substitute expressions (\ref{eqn:pbx}),(\ref{eqn:pby}) into (\ref{eqn:invariance}):}\IEEEnonumber\\
            \lambda_{i\rightarrow{}j}'(t) = \gamma_{i\rightarrow{}j}'(t){+}\beta_{i\rightarrow{}j}'(t)\IEEEnonumber\\ {+}({-}\beta_{i\rightarrow{}j}'(t) {+} \gamma_{i\rightarrow{}j}^{\prime\hspace{3pt}out}(t)){-}(\gamma_{i\rightarrow{}j}'(t) {+} \gamma_{i\rightarrow{}j}^{\prime\hspace{3pt}out}(t))\IEEEnonumber\\
            =\gamma_{i\rightarrow{}j}'(t){+}\beta_{i\rightarrow{}j}'(t) {+}(-\beta_{i\rightarrow{}j}'(t)) {-}(\gamma_{i\rightarrow{}j}'(t))\IEEEnonumber\\
            = 0 \qquad\IEEEQEDhere
        \end{IEEEeqnarray}
    \end{IEEEproof}
\end{lemma}

Due to the inclusion of an initial link occupancy, and the ability to arbitrarily label clocks, $\lambda_{i\rightarrow{}j}$ is not simply the initial placement of frames in the \ac{FIFO} (as it is in the \ac{KPN} model), but also the number of frames in-flight on the link, plus any clock offsets.

\section{Benefits of Elastic Buffers}
\label{sec:analysis}
Both the \ac{LSFP} and bittide implementations are very similar in principle.
Both are Kahn-like token pushing systems with intermediate \ac{FIFO} buffers. The main difference is the process synchronisation behaviour, where \acp{LSFP} use the bounded-\ac{FIFO} model of blocking reads and writes whereas bittide uses a novel clock control system to maintain the bounds of the so-called Elastic \acp{FFP} and hence be blocking free.
\highlight{To highlight the differences between the two proposed implementations of \acp{LSN}, we briefly compare their runtime behaviour via a discrete-event simulator, which is publicly available.\footnote{\url{https://github.com/PRETgroup/bittide_sim_wip}}. We don't simulate non-\ac{LSN} architectures (e.g. \acp{LTTA}), as the structural and paradigmic difference makes the direct comparison difficult. These are designed to be illustrative of the typical execution characteristics of each platform rather than providing a concrete benchmark (as these are only abstract models). The simulation models a list of periodically executing machines, communicating over a set of links with transmission delay. Each machine has a local clock, and a buffer for each incoming link. The simulation progresses by finding the next deadline from either the machine clocks or in-flight messages, and then advancing the simulation time to that point. When the deadline for a machine elapses, a logical tick occurs during which it performs some calculation and updates its next deadline, based on its assigned control policy (e.g. PID).}

\begin{figure}[htp]
    \centering
    \includegraphics[width=0.5\linewidth]{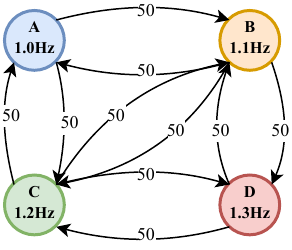}
    \caption{\acs*{LSN} example topology for simulations. The nominal/initial frequency of each machine is annotated.}
    \label{fig:mesh_diagram}
\end{figure}

We consider the \ac{LSN} shown in \Cref{fig:mesh_diagram} implemented over both \ac{LSFP} and bittide. This topology is arbitrarily chosen, but the clock behaviours are indicative for other topologies also.

% \algrenewcommand\algorithmicindent{1.0em}%
% \begin{algorithm}[H]
%     \caption{Top-Level Simulation}
%     \label{alg:system}
%     \begin{algorithmic}[1]
%     \While{simulation not finished}
%         \State Find next deadline $d$ from components
%         \State Progress simulated time $t$ to $d$
%         \ForAll{transmission links}
%             \ForAll{message $m$ with deadline $t_d \leq t$}
%                 \State Call receive(m) on destination buffer
%             \EndFor
%         \EndFor
%         \ForAll{machine M with deadline $d_m \leq t$}
%             \State outputs = $M$.tick()
%             \ForAll{outputs}
%                 \State Send on link with deadline $t + delay$
%             \EndFor
%         \EndFor
%     \EndWhile
%     \end{algorithmic}
% \end{algorithm}
% \begin{algorithm}[H]
%     \caption{Machine Tick Function}
%     \label{alg:machine}
%     \begin{algorithmic}[1]
%     \Procedure{tick()}{}
%         \If{input available}
%             \State Pop from receive buffer
%             \State \textbf{do\_some\_calculation()}
%             \State next\_deadline $\gets$ update\_control() \Comment{e.g. PID}
%             \State \textbf{Return} outputs
%         \Else
%             \State \textbf{do\_skip\_cycle()}
%             \State next\_deadline $\gets$ update\_control()
%             \State \textbf{Return} null
%         \EndIf
%     \EndProcedure
%     \end{algorithmic}
% \end{algorithm}
 
\subsection{Throughput}
\begin{figure}[htp]
    \centering
    \begin{subfigure}[t]{\linewidth}
        \centering
        \includegraphics[width=0.72\linewidth]{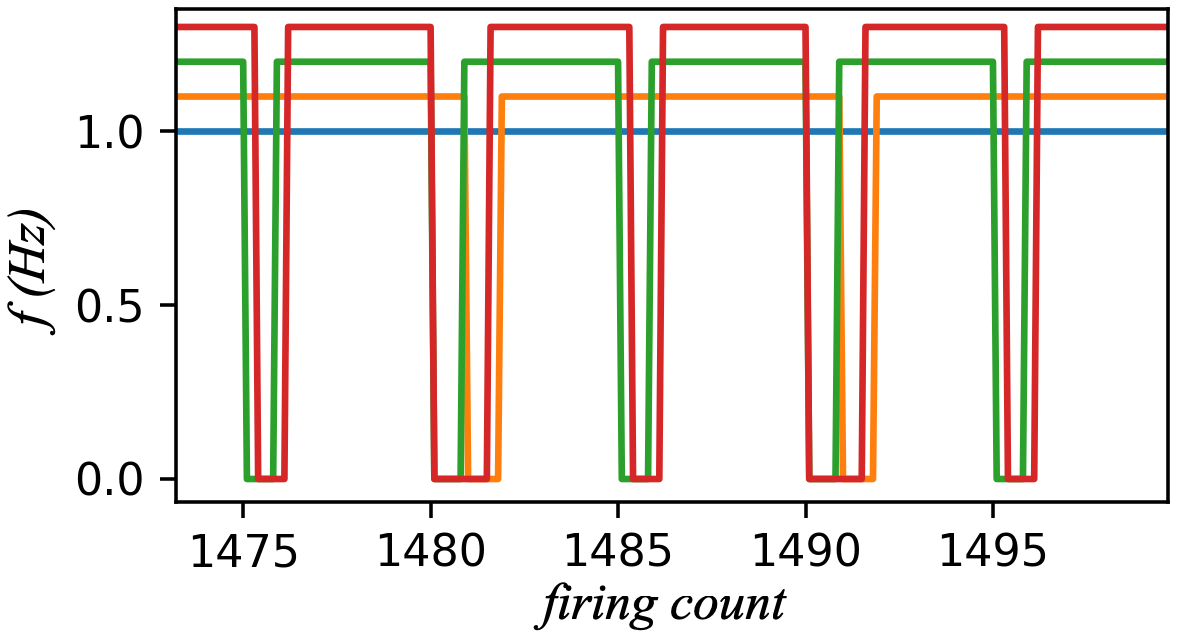}
        \caption{\acs*{LSFP}}
        \label{fig:mesh:ffp}
    \end{subfigure}

    \bigskip

    \begin{subfigure}[t]{\linewidth}
        \centering
        \includegraphics[width=0.72\linewidth]{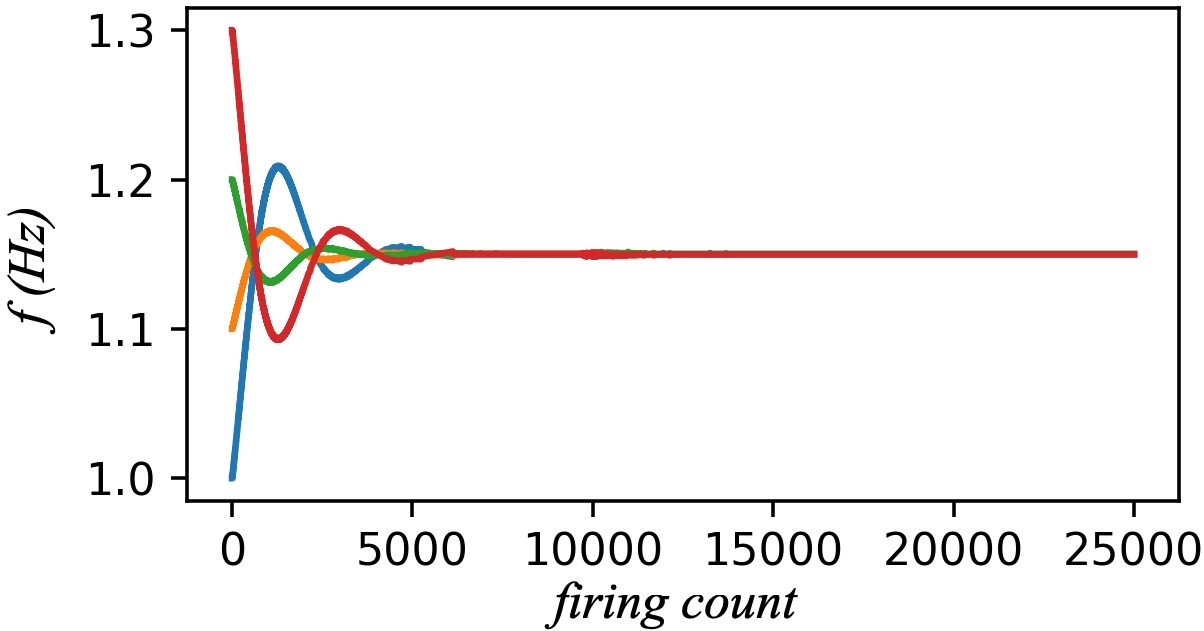}
        \caption{bittide}
        \label{fig:mesh:bittide}
    \end{subfigure}

    \caption{Comparison of \acs*{LSFP} and bittide tick/firing rate on a fully-connected four-machine network. Each coloured line represents a unique machine clock.}
    \label{fig:mesh}
\end{figure}

\highlightg{First, we examine the throughput of each architecture. Throughput here refers to the number of logical ticks that occur within some time period for some given machine, and hence how many calculations are completed. Because our machines are periodic, we consider the frequency of each logical clock as equivalent to the throughput of the machine. }

For our example topology, \Cref{fig:mesh} shows the frequency of each local clock as it evolves during an execution for both \ac{LSFP} and bittide systems. For the \ac{LSFP} implementation (\Cref{fig:mesh:ffp}), the flow of tokens is limited by the slowest machine in the system. The slowest machine can remain free running, while the others must `stutter' to modulate to the same value as the slowest machine on average. 

In contrast, the bittide proportional-integral controller (\Cref{fig:mesh:bittide}) converges all machine frequencies near the midpoint without ever blocking. In systems where task speed-up is possible, the bittide approach may have superior peak throughput. When task rates are very close to begin with, the throughput will be similar for both implementations, but the stabilising effect of elastic buffers will still reduce jitter in the bittide model. Note that logical synchrony is still preserved during the transient or other instability as long as no frame data is lost due to overflow.

\begin{observation}
    Every machine in \iac{LSFP} or bittide network must have the same average firing rate.
    \label{averagethroughput}
\end{observation}

Due to \ac{FIFO} bounds, for each producer-consumer pair the average rate of production must be the same as the average rate of consumption, barring some small initial difference before a steady-state is reached. If rates were unequal, \acp{FIFO} would experience unbounded growth or loss during an extended execution and cause overflow.

\subsection{Physical Latency}
\highlightg{Physical latency between machines refers to the time taken for a token to travel from the producer to the head of the consumer buffer. This is a distinctly different measure of performance to throughput, as latency measures the fastest time a machine can respond to data produced by another machine, rather than how much bulk data is processed.} $\tau_{i\rightarrow{}j}(t)$ denotes the time elapsed between a production at $\theta_i(t){=}s$ and its consumption at $\theta_j(t{+}\tau_{i\rightarrow{}j}(t)){=}s{+}\lambda_{i\rightarrow{j}}$. This is not equivalent to transmission delay $l$ alone, because the token will be held in the recipient buffer for some time.

\begin{observation}
    The latency $\tau_{i\rightarrow{}j}(t)$ between two machines consists of the transmission delay and buffer propagation time
    \begin{equation}
        \tau_{i\rightarrow{}j}(t) \approx l_{i\rightarrow j} + \frac{\beta_{i\rightarrow{j}}(t)}{\omega_j(t)}
        \label{eqn:latencyeqn}
    \end{equation}
\end{observation}

\begin{figure}[htp]
    \centering
    \begin{subfigure}[t]{\linewidth}
        \centering
        \includegraphics[clip,trim=0cm 0cm 1.0cm 0cm, width=0.72\linewidth]{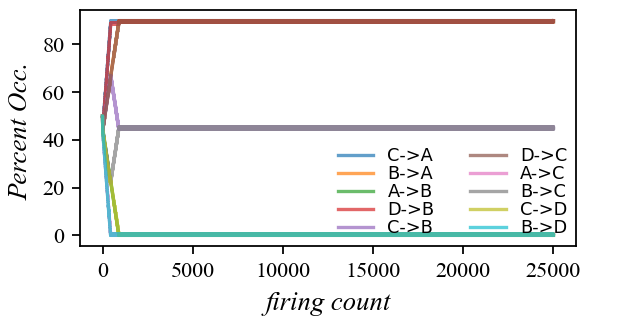}
        \caption{\acs*{LSFP}}
        \label{fig:mesh_lat:ffp}
    \end{subfigure}

    \bigskip

    \begin{subfigure}[t]{\linewidth}
        \centering
        \includegraphics[clip,trim=0cm 0cm 1.0cm 0cm, width=0.72\linewidth]{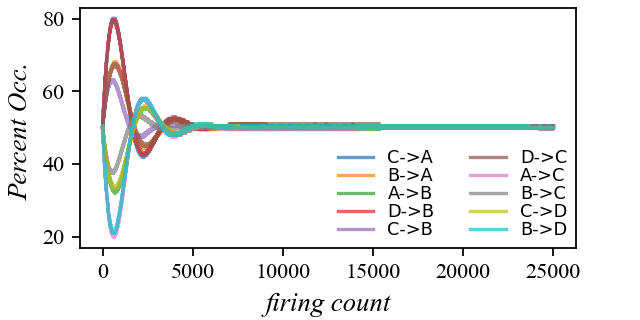}
        \caption{bittide}
        \label{fig:mesh_lat:bittide}
    \end{subfigure}

    \caption{Comparison of \acs*{LSFP} and bittide channel buffer occupancy on a fully-connected four-machine network. Each coloured line represents a single channel.}
    \label{fig:mesh_lat}
\end{figure}

Note the implicit assumption that receiver frequency does not vary much during transmission. Because transmission delay is common to the latency for both implementations, we will compare just the buffer occupancy behaviour of each. \Cref{fig:mesh_lat} shows each communication channel's buffer occupancies for the bittide and \ac{LSFP} example during a slice of execution.

Due to the clock controller, bittide buffers tend towards their initial occupancy. As a result, each bittide channel will (at steady state) have approximately the same latency at
\begin{equation}\tau^{bt}_{i\rightarrow{}j}(t) \approx l_{i\rightarrow j} + \frac{\beta_{i\rightarrow{j}}^{max}}{2\omega_j(t)}\end{equation}

This is because $\beta_{i\rightarrow{j}}(t)$ from \Cref{eqn:latencyeqn} will evaluate to the buffer midpoint $\beta_{i\rightarrow{j}}^{max} / 2$ at steady-state.

In contrast, in \iac{LSFP} a buffer from a slower node to a faster node will tend to be near-empty and a buffer from a faster node to a slower node will tend to fill up before the skipping action kicks in. Note that the fast-to-slow link $D\rightarrow B$ in the simulation has a high occupancy, and the slow-to-fast link $B\rightarrow D$ has an almost-zero occupancy at all times. Consequently, links which end up empty will have shorter latencies than those which end up full :
\begin{equation} l_{i\rightarrow j} \lessapprox \tau^{ffp}_{i\rightarrow{}j}(t) \lessapprox l_{i\rightarrow j} + \frac{\beta_{i\rightarrow{j}}^{max}}{\omega_j(t)}\end{equation}

\highlight{Thus, for two identical systems the elastic buffer approach will tend to have a more equal distribution of latencies across all channels, while the \ac{LSFP} approach will have a more variable distribution. Although, some overhead is incurred by the elastic buffer system, which needs a larger number of initial tokens than \ac{LSFP} for use as control system feedback.}

\subsection{Discussion}
When comparing the two platforms as \ac{LSN} implementations, the more advanced clock control mechanism of the bittide approach tends to improve throughput and make latency more consistent (after a transient period) compared to a blocking \ac{LSFP} approach. \Cref{fig:mesh} and \Cref{fig:mesh_lat} are generated from a specific topology, but the nature of the graphs is invariant to the topology given that a stable bittide controller exists. Thus, we demonstrate the key benefits of the decoupled elastic buffer \ac{MoC} over blocking approaches, for the first time.

This is not entirely surprising given that we implicitly present the assumption in these results that bittide clocks can increase in speed. Even if we remove this assumption and make the control one-sided (not exceeding the initial frequency), then clocks would be expected to settle at the rate of the slowest machine, as is the case in \ac{LSFP}. The benefits to latency and jitter would still be present however, as the system remains blocking-free.

A major redeeming property of \acp{LSFP} is that the governing
blocking mechanism can be implemented over many generic hardware platforms.
In contrast, the dynamic clock control of bittide systems poses a reasonable implementation hurdle,
and we generally assume that it remains a physical-layer protocol for bespoke hardware.
This does not preclude the possibility that bittide-like control could be implemented at a
more abstract software level.
\acresetall
\acuse{FIFO}

\section{Conclusion}
\raggedbottom
\label{sec:conclusion}
This paper addresses a gap of realisable formal models that preserve determinism and free running behaviour in distributed systems.
We formalise a distributed network of machines as \iac{LSN}. \Acp{LSN} provide
a synchronous abstraction of a distributed system.
We show that \acp{KPN}, and their implementation as \acp{LSFP}, preserve the \ac{LSN} model.
Subsequently, we show that the bittide protocol for
distributed systems also preserves the \ac{LSN} model. Thus, \acp{LSFP} and bittide offer two distinct ways of realising
\ac{LSN} behaviour. We observe that in both \acp{LSFP} and bittide,
the Kahn-reminiscent effect of pipelining via buffers is used to achieve a greater overall throughput compared to
signalling approaches used in \ac{GALS} models, while preserving determinism,
which is difficult to achieve over \ac{GALS}. We also show that the decoupling of execution from synchronisation in bittide leads to higher performance.

While this paper paves the way for formalising distributed synchronous systems using \acp{LSN}, we have several avenues for future research.
Firstly, we need to consider the design of application models which leverage logical synchrony. There is scope for introducing novel constructs into synchronous languages which enable the seamless distribution of programs over bittide-like platforms. Moreover, precise delays expressed in logical-time have implications for efficient dataflow scheduling.

\section*{Acknowledgments}
This work was supported in part by a generous grant from Google Research entitled
``Designing Scalable Synchronous Applications over Google bittide''. We give special
thanks to other members of the bittide team, including but not limited to
Pouya Dormiani, Chris Pearce,
and Robert O'Callahan for their continuous feedback and support.
Kenwright and Roop also acknowledge constructive feedback received
from Michael Mendler and Gerald L\"{u}ttegen from Bamberg University. Sanjiva Prasad of IIT Delhi provided significant editorial feedback contributing to the revised manuscript. 

\bibliographystyle{IEEEtran}
\bibliography{refs1204}
\clearpage

\begin{IEEEbiography}[{\includegraphics[width=1in,height=1.25in,clip,keepaspectratio,trim=2.5cm 2.5cm 2.5cm 5cm]{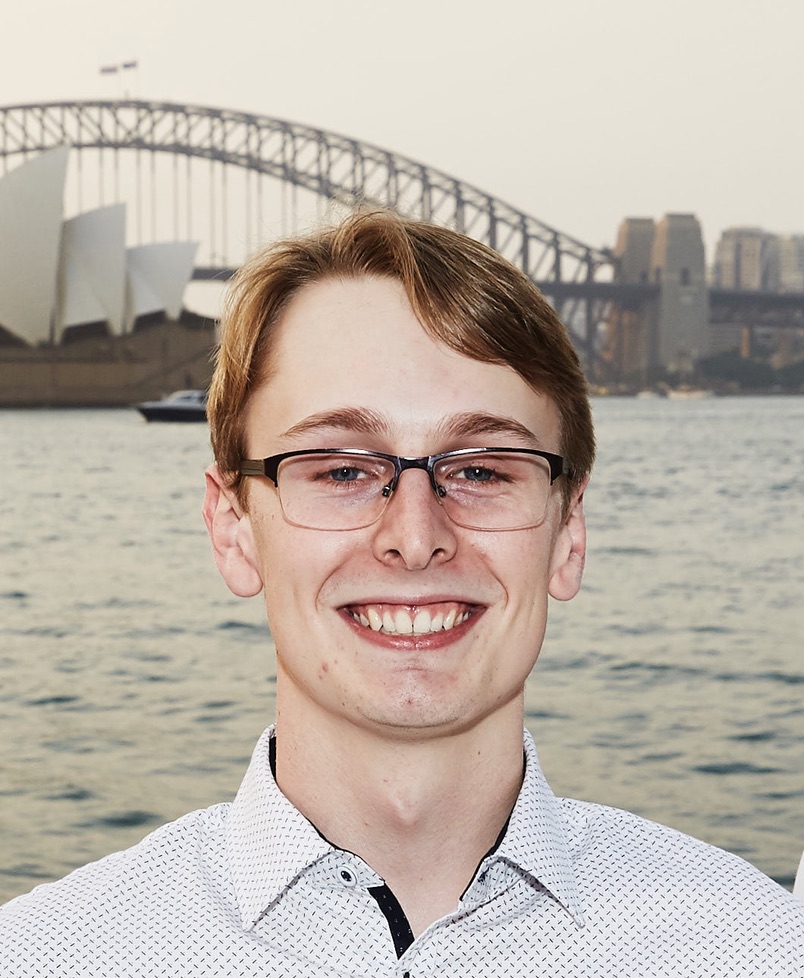}}]{Logan Kenwright} is a PhD student under Partha Roop at the University of Auckland. He holds a BE degree in Computer Systems Engineering with first-class honours from the University of Auckland. Logan's research interests include programming models for synchronous and deterministic systems.
\end{IEEEbiography}

\begin{IEEEbiography}[{\includegraphics[width=1in,height=1.25in,clip,keepaspectratio,trim=1cm 6cm 1cm 2cm]{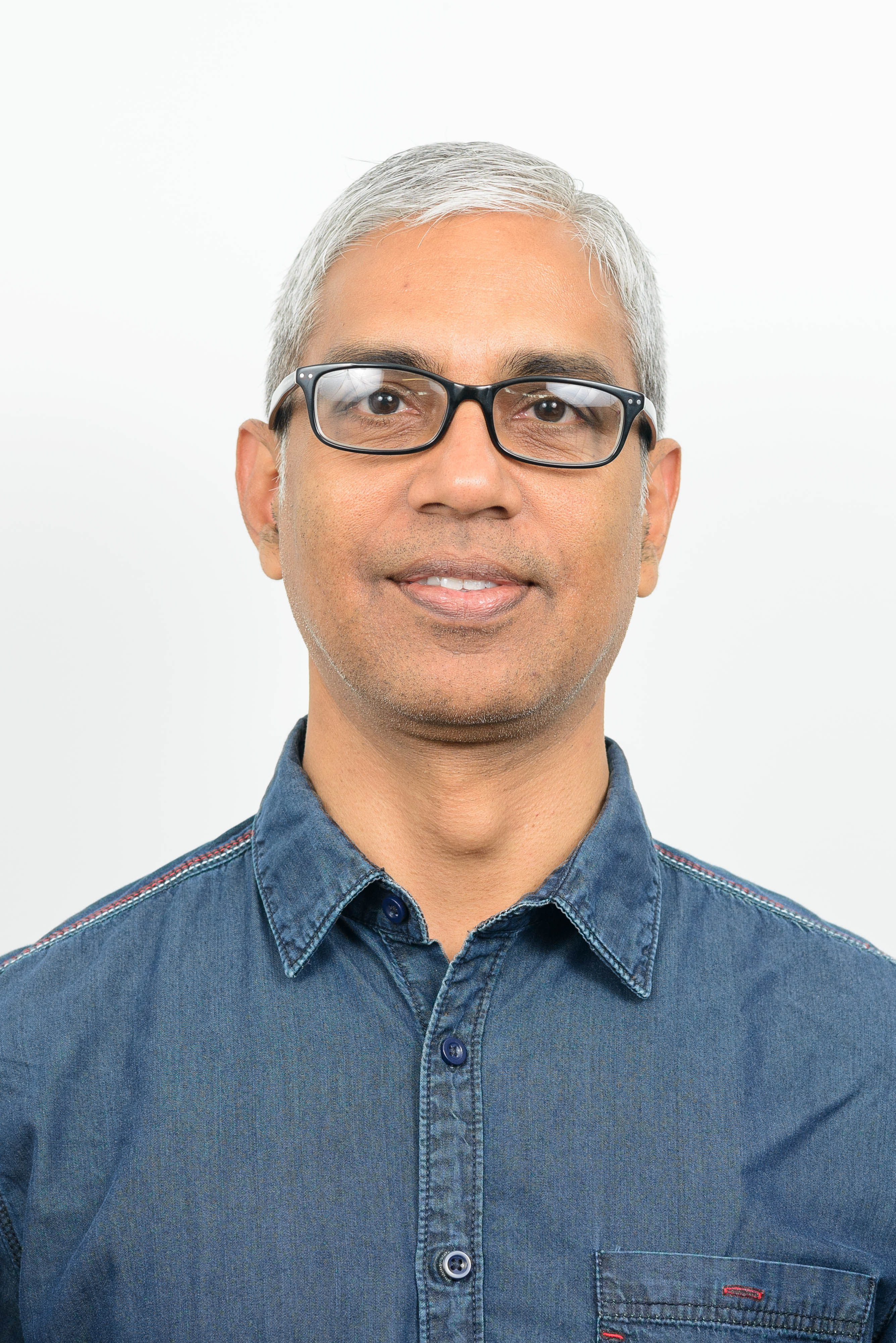}}]{Partha Roop} is a Professor and Associate Dean International of the Faculty of Engineering, The University of Auckland. Partha received a PhD from UNSW (Sydney), M.Tech from IIT, Kharagpur and B.E from Anna University(CEG, Guindy), all in computer science and engineering. Partha's research interests are Formal Methods for Safety-Critical Software,  AI and Machine Learning especially focussing on safety, and Real-Time Systems. He is a steering committee member of IEEE/ACM International Conference on Formal Methods and Models of Codesign (MEMOCOE) and an Associate Editor of IEEE Embedded Systems Letters. Recently, Partha is involved in the global partnership on AI for future pandemic resilience (\href{https://gpai.ai/}{gpai.ai}).
\end{IEEEbiography}

\begin{IEEEbiography}[{\includegraphics[width=1in,height=1.25in,clip,keepaspectratio]{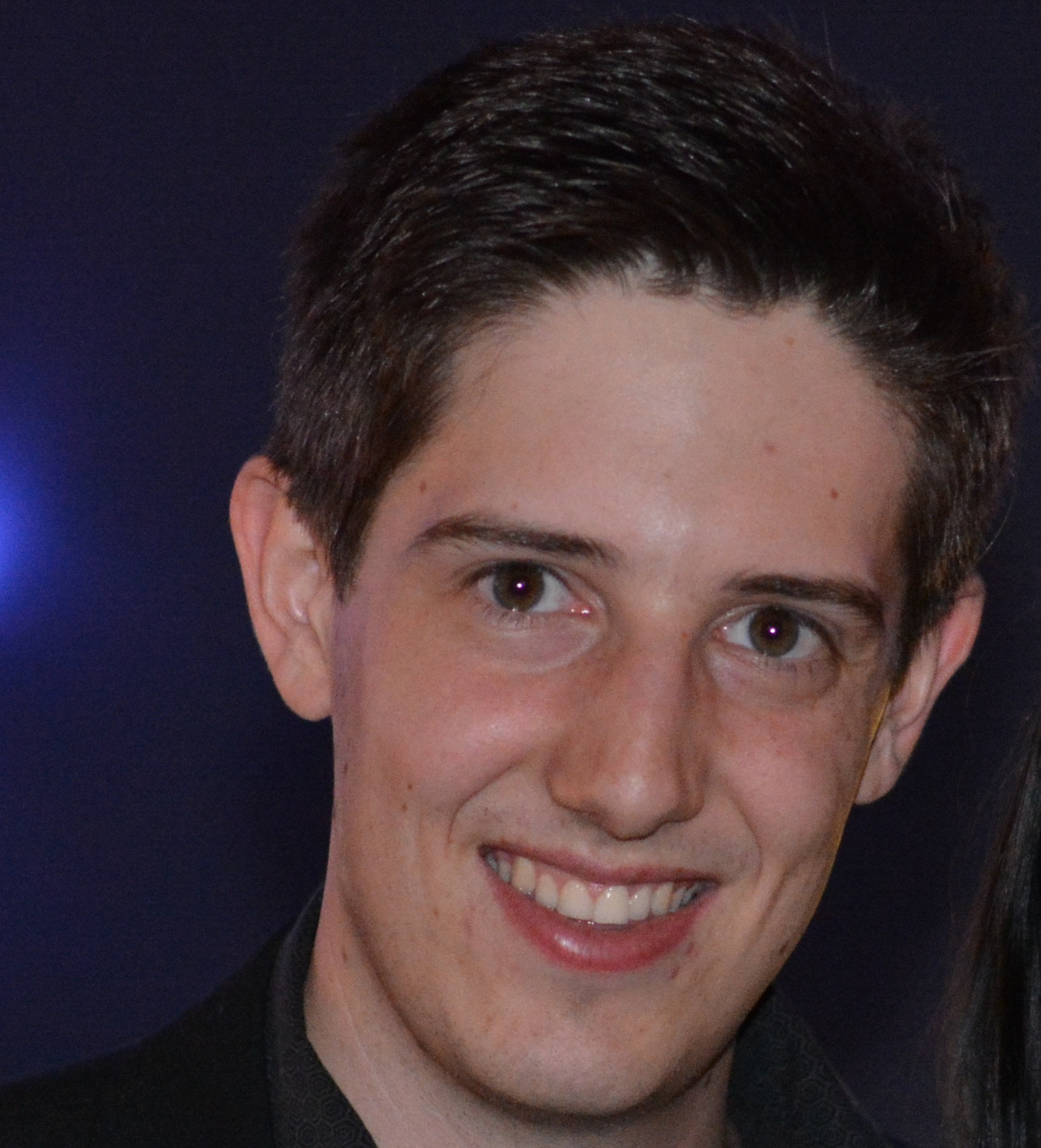}}]{Nathan Allen}
    is a Research Fellow in the Department of Electrical, Computer, and Software Engineering at The University of Auckland.
    He received a B.E. (Hons) degree in Computer Systems Engineering in 2015, and a Ph.D. in Computer Systems Engineering in 2021, both from the University of Auckland, New Zealand.
    His research interests are in synchronous programming languages, formal methods, and embedded systems, which were all applied to the design of biomedical embedded systems in his Ph.D. research, and subsequently to the use of Spiking Neural Networks in robotic applications.
\end{IEEEbiography}

\begin{IEEEbiography}[{\includegraphics[width=1in,height=1.25in,clip,keepaspectratio]{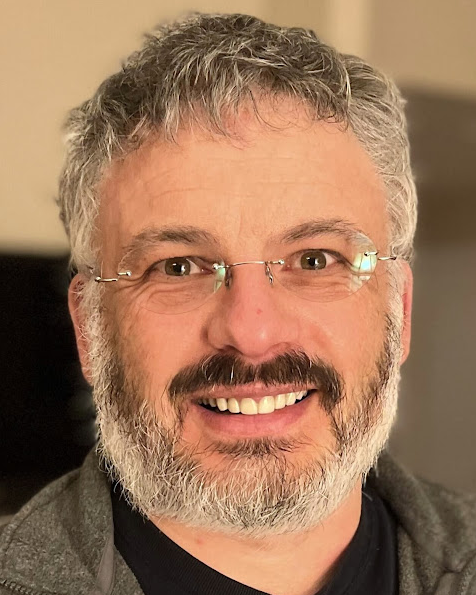}}]{C\u{a}lin Ca\c{s}caval} (Fellow, IEEE) is Director of
    Engineering at Google Research, leading research in scalable distributed
    systems and compilers. He holds MS degrees from Technical University of Cluj,
    Romania and West Virginia University, USA, and in 2000 received a Ph. D. degree
    in Computer Science from University of Illinois at Urbana-Champaign. C\u{a}lin
    spent his career in industrial research, where identified industry trends,
    defined, built, and delivered first of a kind prototypes and products,
    including: the first programmable networking (P4) production compiler and
    networking stack at Barefoot Networks; the first mobile heterogeneous
    computing runtime and parallel browser, mobile optimized math libraries and
    power optimization framework at Qualcomm Research; system software for the
    Blue Gene family of supercomputers and the first UPC compiler to scale to
    hundreds of thousands of processors at the IBM TJ Watson Research Center.
\end{IEEEbiography}

\begin{IEEEbiography}[{\includegraphics[width=1in,height=1.25in,clip,keepaspectratio]{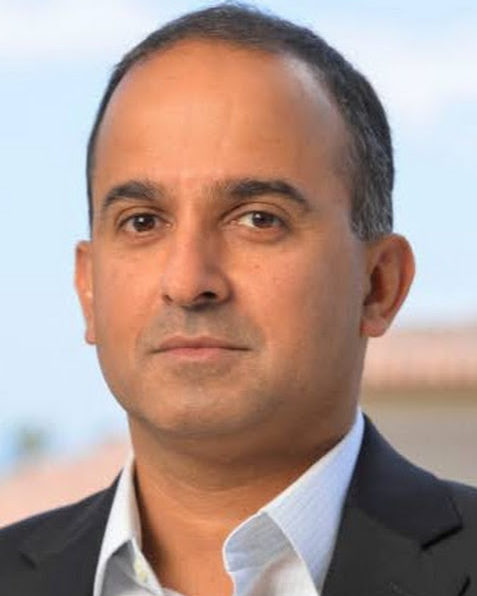}}]{Sanjay Lall} (Fellow, IEEE)
    is Professor of Electrical Engineering in the Information
    Systems Laboratory at Stanford University. He received a B.A. degree
    in Mathematics with first-class honors in 1990 and a Ph.D. degree in
    Engineering in 1995, both from the University of Cambridge,
    England. His research group focuses on algorithms for control,
    optimization, and machine learning. From 2018 to 2019 he was Director
    in the Autonomous Systems Group at Apple. Before joining Stanford he
    was a Research Fellow at the California Institute of Technology in the
    Department of Control and Dynamical Systems, and prior to that he was
    a NATO Research Fellow at Massachusetts Institute of Technology, in
    the Laboratory for Information and Decision Systems. He was also a
    visiting scholar at Lund Institute of Technology in the Department of
    Automatic Control. He has significant industrial experience applying
    advanced algorithms to problems including satellite systems, advanced
    audio systems, Formula 1 racing, the America's cup, cloud services
    monitoring, and integrated circuit diagnostic systems. He is currently
    a visiting researcher at Google.
\end{IEEEbiography}

\begin{IEEEbiography}[{\includegraphics[width=1in,height=1.25in,clip,keepaspectratio]{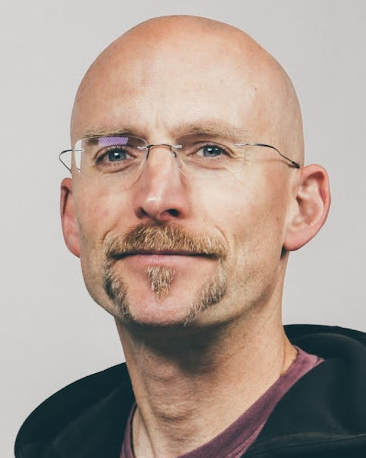}}]{Tammo Spalink}
    was born in Germany but mostly grew up in the USA.  He holds a BS
    degree from Carnegie Mellon University, an MS degree from the
    University of Arizona, and a PhD from Princeton University -- all in
    Computer Science.  Tammo has spent his career to date at Alphabet
    where he has contributed to Android, ChromeOS, Loon, and numerous
    internal projects.  Tammo is currently an Engineering Director in
    Google Research responsible for a range of projects from silicon
    design to machine learning compilation.
\end{IEEEbiography}

\begin{IEEEbiography}[{\includegraphics[width=1in,height=1.25in,clip,keepaspectratio]{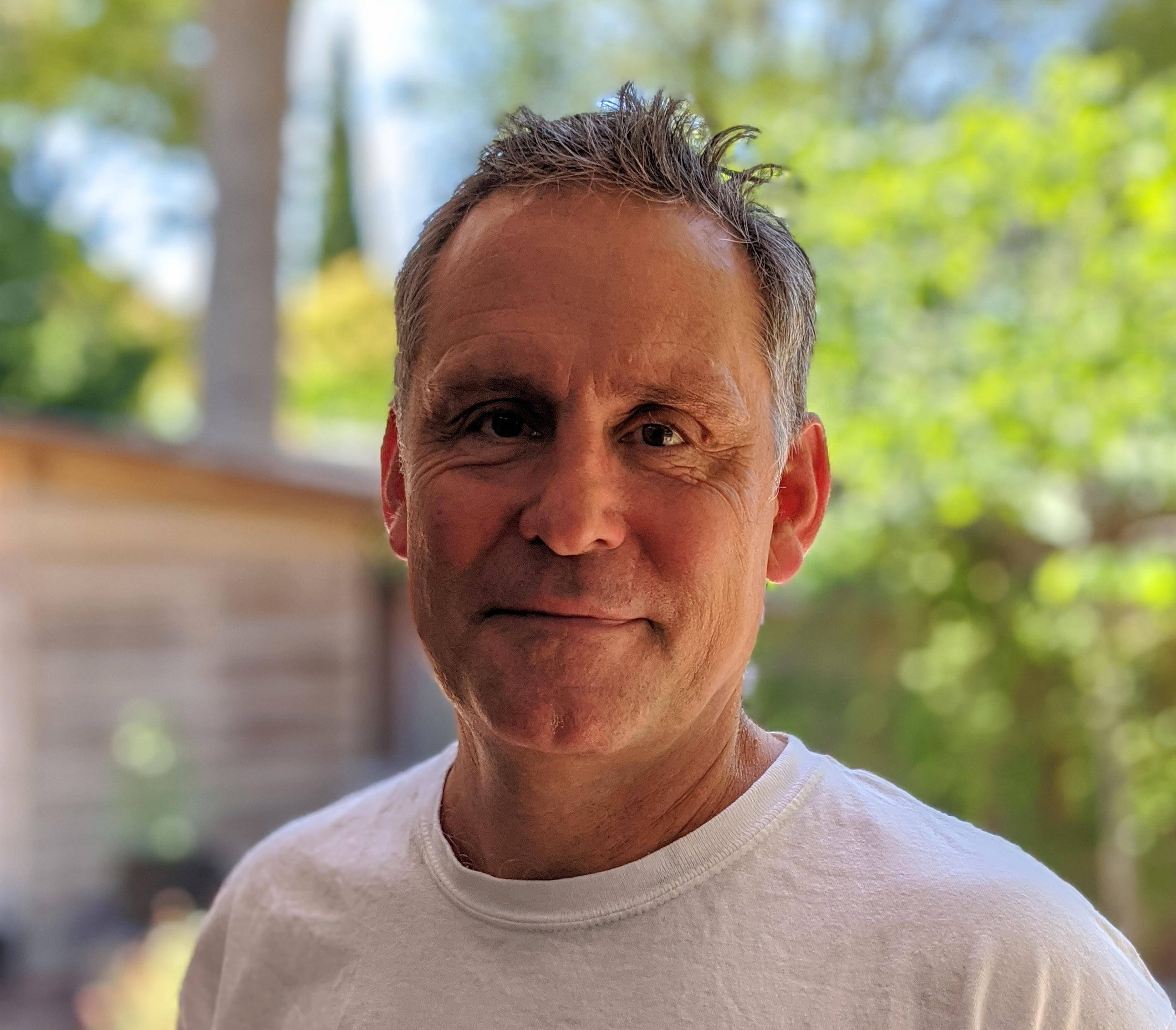}}]{Martin Izzard}
    grew up in Durban, South Africa and was awarded BSEE and MSEE degrees from Natal University (today the University of KwaZulu-Natal).  He completed a PhD at Trinity College Cambridge and was also awarded a Title A Fellowship at Trinity. He was at Texas Instruments in Dallas Texas for 21 years, starting in research and then holding a variety of technical- and general-management roles in both the digital and analog divisions of TI. He co-founded an Ethernet Switch chip company in 2017. He joined Google in 2019 as a research director.
\end{IEEEbiography}

\EOD

\end{document}